# Play Me Something "Icy": Practical Challenges, Explainability and the Semantic Gap in Generative AI Music

**Jesse Allison**
Louisiana State University
Baton Rouge, USA
jtallison@lsu.edu

**Drew Farrar**
Louisiana State University
Baton Rouge, USA
afarra6@lsu.edu

**Treya Nash**
Louisiana State University
Baton Rouge, USA
tnash6@lsu.edu

**Carlos G. Román**
Louisiana State University
Baton Rouge, USA
cromcn1@lsu.edu

**Morgan Weeks**
Louisiana State University
Baton Rouge, USA
mweeks8@lsu.edu

**Fiona Xue Ju**
Louisiana State University
Baton Rouge, USA
xju1@lsu.edu

**ABSTRACT**

This pictorial aims to critically consider the nature of text-to-audio and text-to-music generative tools in the context of explainable AI. As a group of experimental musicians and researchers, we are enthusiastic about the creative potential of these tools and have sought to understand and evaluate them from perspectives of prompt creation, control, usability, understandability, explainability of the AI process, and overall aesthetic effectiveness of the results. One of the challenges we have identified that is not explicitly addressed by these tools is the inherent semantic gap in using text-based tools to describe something as abstract as music. Other gaps include explainability vs. useability, and user control and input vs. the human creative process. The aim of this pictorial is to raise questions for discussion and make a few general suggestions on the kinds of improvements we would like to see in generative AI music tools.

**Author Keywords**

generative AI; prompts; text-to-sound; text-to-music; semantic gap; experimental music

**CSS Concepts**

• Human-centered computing~Human-Computer interaction

## INTRODUCTION

As a group of experimental musicians, we are interested in exploring cutting-edge digital tools and technologies that have the potential to enhance music composition, performance, and production. The recent development of text-to-sound AI tools offer many possibilities for artistic creation, but also raises many questions regarding the aesthetic and ethical implications of their usage. Through evaluating prominent, publicly available text-to-audio and text-to-music generative AI tools [1, 2, 3, 4, 5], we seek to understand and compare their general implications from various perspectives, including prompt creating, control, usability, understandability, explainability of the AI process, and overall effectiveness of the results.

In order to respond to these tools, each member of our group experimented individually, exploring different prompt styles and inputs, taking into account the degree of explainability in the processes used to generate the audio. To ascertain the versatility of the tools, we varied prompts based on musical semantics, descriptors, complexity, genre, and musical style. The artifacts produced during these explorations were the basis for the insights presented in this pictorial. Following our individual experiments, we held regular meetings to share our observations. These sessions allowed for in-depth discussions on the general approach of these kinds of technologies, including their strengths, weaknesses, and potential applications. During our meetings, we employed design thinking methodologies such as sketching and mental mapping to structure our ideas. Through our discussions and reflections, we identified an overarching theme of the semantic gap between music and language within the context of generative AI music tools.

## THE GAP: SEMANTIC AND OTHERWISE

When generating artistic material, AI models must deal with the technical knowledge and specialized vocabulary associated with various artistic domains. In music generation, familiarity with music theory, instrumentation, sound descriptors and genre conventions is indispensable. Unlike other forms of data, such as text or images, many musical elements defy precise and well-defined description. This causes a 'semantic gap' [6]. The semantic gap poses a significant challenge for generative AI music tools in dealing with the subjective nature of musical description and expression. This gap is expanded by culture-specific musical descriptions. Prompt inputs have become one of the most widely used methods of interfacing with generative AI tools [7]. They allow users to interface with AI tools without needing to understand their inner workings. The semantic gap between language and music means that linguistic prompts may fail to capture the abstractness of musical ideas. While prompt refinement offers a degree of control, the distinction between "good" and "bad" prompts in music creation remains ambiguous. Text-to-audio and text-to-music tools face the challenge of capturing musical ideas, translating them into words, and translating them back into coherent sonic material.

An interesting prospect lies in the ability to fine-tune AI sound generation systems to the users' own sets of descriptors, categories, or tags. By aligning the system's output with users' unique preferences and vocabularies, generative AI tools could offer a more personalized and intuitive creative experience. Explaining the relationship between user descriptors and AI responses could bring greater transparency and interpretability in the generative process.

When AI tools clearly communicate their functionality and capabilities, users can more precisely control parameters and output.

If AI tools require too much pre-existing knowledge, this may limit potential creative output, especially for beginner users/musicians.

Some generative AI music tools seem geared towards producing finished products quickly with little input or control. This kind of practice can be well suited for beginner users or commercial areas where fast content generation is valued.

Lack of control can frustrate musicians when outputs don't align with their artistic vision. This unpredictability may also lead to irrelevant results, requiring extra effort to sift through generated content for usable material.

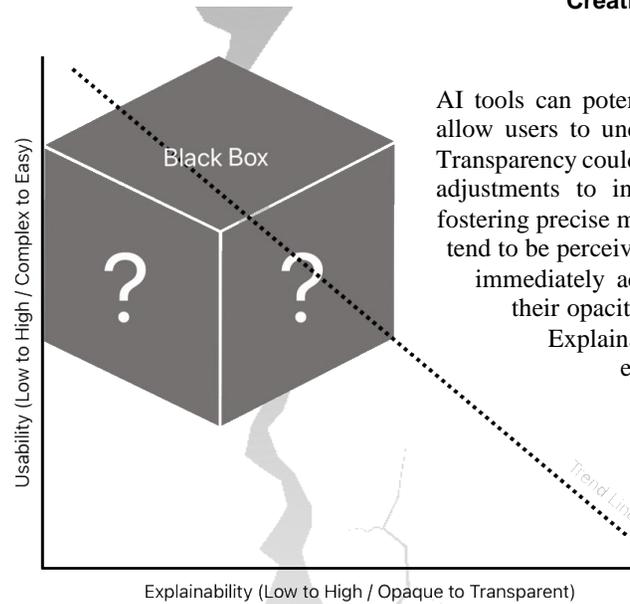

AI tools can potentially become more creatively useful when they allow users to understand the processes that lead to their outputs. Transparency could enable musicians the freedom to make meaningful adjustments to inputs that alignment with their artistic visions, fostering precise musical expression. AI tools with obscure processes tend to be perceived as 'black boxes', which, while potentially more immediately accessible, may alienate professional users due to their opacity. Bryan-Kinns highlights the importance of using Explainable AI (XAI) to bridge the gap between an artist's expectations and the AI's output, thereby enhancing the tool's utility for professional creative applications [8].

Generative AI music tools, especially those driven by text-to-audio models, often operate through simple text inputs on web interfaces without offering clear explanations on how the inputs are transformed into musical outputs. Musicians often rely on discrete musical descriptors (pitch, rhythm, tempo, etc.) to convey information. Tools that communicate correlations using the aforementioned musical language could allow users to align outputs with their creative intentions. While tools may not need to explain their process in detail, transparency in training data and intended use cases could lead to more obvious cause and effect, and thus usability.

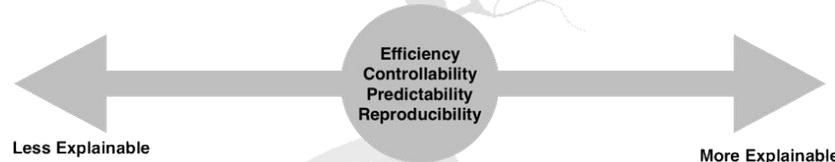

## Prompt:

Line drawing of a domestic shorthair cat slinking around

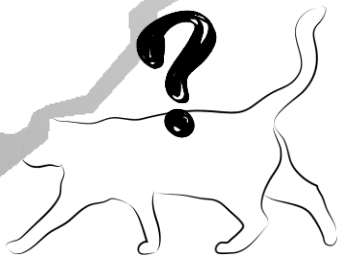

## Outputs:

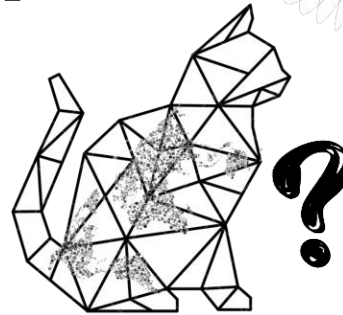

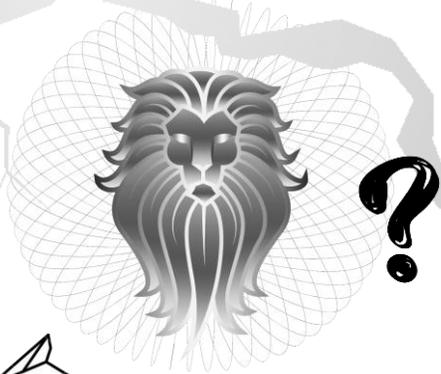

### Success or Failure
### Error or Creativity

What defines success? An exact reproduction of the prompt or a whimsical interpretation? Does introducing error increase creative potential or destroy usable results?

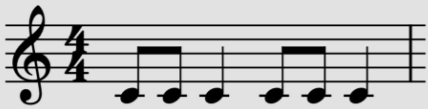
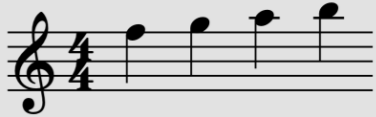
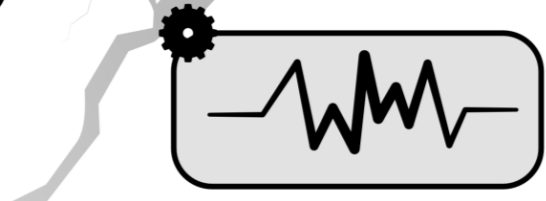
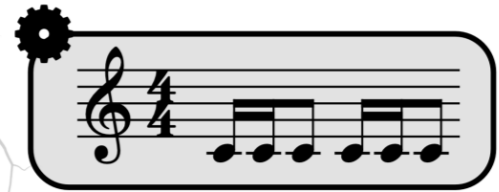
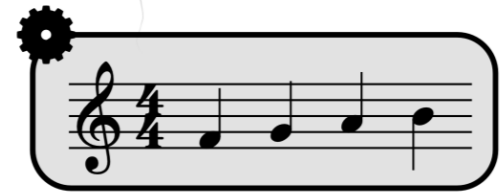

The gap between the user control of generative AI music tools and their input methods reduces their potential for practical use. Many generative AI music tools used text-based prompts to generate full or partial songs. Currently, song generation tools do not allow for the isolation and adjustment of specific parameters. For example, as represented in the visual above, when the user inputs "a dark, icy section" into a generative AI music tool the output could be in the form of a pop song with a winter theme. What part of the input said that the result should be a pop song? Moreover, what is "icy" in music? Every user will have a different idea of how "icy" translates to music. In these situations, the user must regenerate the music using a slightly altered prompt, which may lead to a significantly different outcome without allowing them to make adjustments to the previously generated music. As some have discussed, the rigidity of current AI generative content (AIGC) tools does not mesh well with the iterative, trial-and-error creative process typical of many artists [9, 10, 11].

In contrast, Computer Assisted Composition (CAC) tools, which are used to augment common musical creation processes, tend to use interfaces that are familiar to musicians. Software such as MAX, OpenMusic, and Bach incorporate a user interface that includes traditional Western notation and visual programming, allowing the user to manipulate the music directly while also producing outputs that are more malleable and familiar (often standard musical notation). These tools lessen the gap between the user and the desired output in comparison to the current generative AI music tools. However, these tools require an additional layer of expertise from users, creating a barrier for even some experts within the field. Traditional compositional techniques and CAC environments can provide a model for AI tools to structure more useful interfaces with a more usable output, lessening the user-to-output gap while reducing the specific expertise required.

The output of these tools being partial or full-length songs, along with their conflict with the traditional creative process, raises questions about who, and what, these tools are intended for. These questions lead to the thorny and widely discussed issue of ethics in AI. Makayla Lewis discusses the "transparency of attribution," proposing discussion questions about where AI gets its training data, "who are the contributors? Why were they selected? Were permissions obtained? and does this lack of transparency impact trust in the responses and or negatively impact the arts community?" [12] Most of the music tools we looked at do not reveal where their datasets come from. How can users trust these tools amidst the fear that their creative property will be stolen and remade into generic commercial reproductions? Transparency in things like where the data comes from, how it is being obtained, the size of datasets, and how the data is being labeled would allow generative music tools to become more usable, explainable, and trustable, leaving room for experimentation without the threat of plagiarism, theft, or unlicensed reproduction. A fundamental disconnect (or gap) divides opaque training and classification processes and the human artistic process. This gap can begin to be broached through openness about where — and who — the data is coming from.

## CONCLUSION

As we conclude our exploration of practical issues in generative AI tools for sound and music creation, it becomes evident that certain challenges, such as the semantic gap, are unlikely to be fully resolved in a text-to-sound generative model. Moreover, while text-to-sound and prompt engineering approaches offer some degree of intuitive control over the creative process, they may not represent the optimal paradigm for interacting with a generative music tool.

That these tools are designed to create full songs generated from scratch raises questions about whether this kind of outcome can align with a human-centered approach to AI.

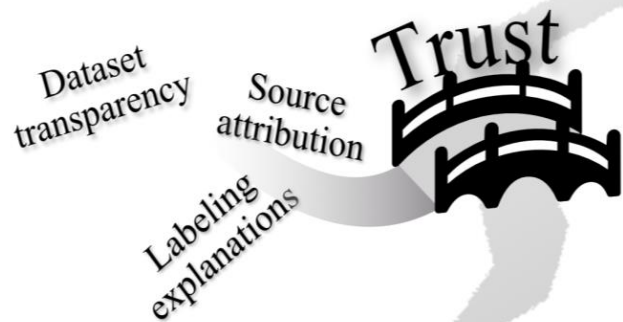
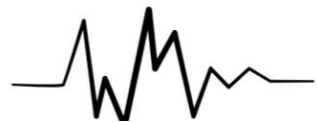

As AI systems increasingly make decisions traditionally reserved for human creators, concerns arise regarding the ethical and aesthetic implications of delegating creative agency to algorithms. Against this background the threat of AI tools replacing human creation, especially in commercial spheres, continues to grow.

Without transparency regarding the sources from which the generative AI tools are being trained, concerns about plagiarism and artistic integrity remain unresolved.

Some of the improvements we would like to see in these tools include clearer articulation of their methodologies and limitations, user interfaces that allow musicians to draw on their expertise, and, perhaps most importantly, transparency in where and who the training data is coming from.

We hope that this pictorial will contribute to the discourse of explainable and human-centered AI in the arts, as well as leading to further discussion, collaboration, and creation.